\documentclass[aps,amsmath,amssymb,notitlepage,twocolumn,nofootinbib,floatfix]{revtex4-1}

\pdfoutput=1

\usepackage[T1]{fontenc}  % Ensure you apt-get install cm-super

\usepackage[version=3]{mhchem} % Package for chemical equation typesetting
\usepackage{graphicx} % Required for the inclusion of images
\usepackage{natbib} % Required to change bibliography style to APA

\usepackage{amsmath} % Required for some math elements
\usepackage{amssymb} % Required for some math elements

\usepackage{hyperref} % Generates links for figures, citations, etc.
\usepackage[all]{hypcap}    %for going to the top of an image when a figure reference is clicked
\usepackage{pgf} % Required for PGF figures
\usepackage{array}
\usepackage[top=1in, bottom=1in, left=1in, right=1in]{geometry}
\usepackage{multirow} % For tables with cells spanning multiple rows.
\usepackage{siunitx}

\usepackage{textcomp}  % Required for \textmu

\usepackage{lipsum}
\newcommand{\us}{\textmu{}s}
\newcommand{\reg}{\textsuperscript{\textregistered}}
\newcommand{\dwave}{\mbox{D-Wave}}  % Do not break word.

\newcommand{\hiderefname}[1]{#1}  % Show author names if they are not shown by the cite command.

\newlength\figureheight 
\newlength\figurewidth

\DeclareGraphicsExtensions{.pdf,.png,.jpg}
\graphicspath{{figures/}}

\newcommand{\hide}[1]{}

\renewcommand{\epsilon}{\varepsilon}%better epsilon

\newcommand{\specialcell}[2][c]{\begin{tabular}[#1]{@{}c@{}}#2\end{tabular}}

\definecolor{titlecolor}{RGB}{0,51,102}
\hypersetup{
	colorlinks = true,
	allcolors = {titlecolor}
}

\begin{document}
	
\title{Quantum Annealing amid Local Ruggedness and Global Frustration}
\author{James King}\email[]{jking@dwavesys.com}
\author{Sheir Yarkoni}
\author{Jack Raymond}
\author{Isil Ozfidan}
\author{Andrew D. King}
\author{Mayssam~Mohammadi~Nevisi}
\author{Jeremy P. Hilton}
\author{Catherine C. McGeoch}
\affiliation{\dwave{} Systems}
\date{\today}

\begin{abstract}
	A recent Google study [{\it Phys.~Rev.~X}, {\bf 6}:031015 (2016)] compared a \dwave{} 2X quantum processing unit (QPU) to two classical Monte Carlo algorithms: simulated annealing (SA) and quantum Monte Carlo (QMC).  The study showed the \dwave{} 2X to be up to 100 million times faster than the classical algorithms.  The Google inputs are designed to demonstrate the value of collective multiqubit tunneling, a resource that is available to \dwave{} QPUs but not to simulated annealing.  But the computational hardness in these inputs is highly localized in gadgets, with only a small amount of complexity coming from global interactions, meaning that the relevance to real-world problems is limited.
	
	In this study we provide a new synthetic problem class that addresses the limitations of the Google inputs while retaining their strengths.  We use simple clusters instead of more complex gadgets and more emphasis is placed on creating computational hardness through frustrated global interactions like those seen in interesting real-world inputs.  The logical spin-glass backbones used to generate these inputs are planar Ising models without fields and the problems can therefore be solved in polynomial time [{\it J.~Phys.~A}, {\bf 15}:10 (1982)].  However, for general heuristic algorithms that are unaware of the planted problem class, the frustration creates meaningful difficulty in a controlled environment ideal for study.

	We use these inputs to evaluate the new 2000-qubit \dwave{} QPU.  We include the HFS algorithm---the best performer in a broader analysis of Google inputs---and we include state-of-the-art GPU implementations of SA and QMC.  The \dwave{} QPU solidly outperforms the software solvers: when we consider pure annealing time (computation time), the \dwave{} QPU reaches ground states up to 2600 times faster than the competition.  In the task of zero-temperature Boltzmann sampling from challenging multimodal inputs, the \dwave{} QPU holds a similar advantage and does not see significant performance degradation due to quantum sampling bias.
\end{abstract}

\maketitle

% disable subsections and subsubsections in the TOC
\makeatletter
\def\l@subsubsection#1#2{}  % disable subsubsections in the TOC
\makeatother

\tableofcontents

\section{Introduction}

Quantum annealers are designed to take advantage of quantum tunneling to find good solutions to hard optimization problems.  When constructing a family of synthetic inputs to test the potential of a quantum annealing platform, one should therefore ensure that the inputs a) are such that solvers can benefit from quantum tunneling, and b) are hard optimization problems with global frustration.

For a solver to benefit from quantum tunneling, the energy landscape associated with the input must have tall, thin energy barriers.  For an input to be computationally hard, the input must have constraints that interact with each other in nontrivial ways.

Quantum processing units (QPUs) developed by \dwave{} Systems
%\footnote{\dwave{}, \dwave{} One, \dwave{} Two, and \dwave{} 2X are trademarks of \dwave{} Systems Inc.} 
that use the quantum annealing algorithm have been commercially available since 2011.  These QPUs solve Ising model inputs defined on the underlying working graph of the chip.  There have been various efforts to evaluate the performance of the \dwave{} systems using synthetic inputs generated randomly from different distributions, or \emph{input classes}.

This study has two main contributions: to propose a new problem class ideal for evaluating D-Wave QPUs, and to use this problem class to evaluate the 2000-qubit D-Wave QPU.

\subsection{Proposing a new problem class}

Previous evaluations of D-Wave QPUs have used problem classes that benefit either too little or too much from quantum tunneling to be ideal for evaluating quantum annealers.  

On one side of this spectrum we have problems such as random unstructured $\pm 1$ problems on the Chimera topology native to D-Wave QPUs.  These were used by R{{\o}}nnow et al.~\cite{Ronnow2014} in their evaluation of the D-Wave Two QPU in 2014, but they are now known \cite{Katzgraber2014} to lack a finite-temperature phase transition, meaning that quantum tunneling is unlikely to play a significant role when solving them.

On the other side of the spectrum, Denchev et al.~\cite{Denchev2016} recently introduced an input class designed to benefit immensely from quantum tunneling.  We refer to these inputs as Google problems.  Their study showed a massive speed increase (up to 100 million times faster) of a \dwave{} 2X system over simulated annealing (SA) and quantum Monte Carlo (QMC), also known as \emph{simulated quantum annealing}.  This provided strong evidence for the ability of quantum annealing to leverage quantum tunneling in a computationally relevant way.  However, the spin-glass backbones of the Google problems are easy to solve, meaning that a) the problems have limited relevance to real-world problems, and b) certain cluster-detecting algorithms can solve them with relative ease \cite{Mandra2016}.

In this study we provide a problem class that aims to retain the advantages of Google problems while being more reflective of real-world problems.  They are more reflective of real-world problems because, rather than relying too heavily on finely-tuned gadgets, they derive much of their computational hardness from larger spin-glass backbones with planted frustration.

Our problems are synthetic and are easy to solve using knowledge of the problem class.\footnote{For example, the super-spin heuristic \cite{Mandra2016} that relies on hard-coded knowledge of clusters would be far faster than the software solvers we consider.  However, such heuristics do not generalize to other problem classes and it would not make sense to include them as competition solvers.}  More specifically, since the logical problems are Ising models on a planar lattice without fields, they are solvable in polynomial time \cite{barahona1982computational}.  However, frustration in the logical problems creates meaningful difficulty for heuristic methods that are unaware of the planted problem class.  Further, the inputs have properties such as tunable ruggedness that make them useful for the evaluation of quantum annealing and classical approximations thereof.  In this way, they are similar to Kauffman's NK model that has proved very useful in the analysis of evolutionary algorithms \cite{Kauffman1987,Kauffman1989,Weinberger1996}.

\subsection{Evaluation of the 2000-qubit D-Wave QPU}

We use this new problem class to evaluate the latest-generation D-Wave QPU.  We measure its performance in absolute terms and we analyze its response to the ruggedness parameters of the problem class.

The software competition we consider is much stronger than that considered by Denchev et al.~\cite{Denchev2016}, and includes GPU implementations of SA, QMC, and SVMC, and also includes Selby's implementation \cite{Selby2014,Selby2013git} of the Hamze-de Freitas-Selby (HFS) algorithm \cite{Hamze2004,Selby2014}.  In the study of Mandr\`{a} et al.~\cite{Mandra2016} that used a wide array of algorithms to solve Google problems, Selby's implementation of HFS was the fastest software solver in terms of both scaling and absolute speed.

We find that the \dwave{} QPU is able to find ground states up to 2600 times faster than the software competition.  We also consider the problem of sampling from ground states and find that the \dwave{} QPU maintains a similar advantage and does not struggle to find a diverse set of optimal solutions.

The remainder of the paper is organized as follows.  In Section \ref{sec:dwave} we provide a description of the 2000-qubit \dwave{} system and a history of \dwave{} QPUs.   In Section \ref{sec:fcl} we present the problem class analyzed in this paper and discuss the concept of ruggedness and its relevance to optimization problems.  In Section \ref{sec:solvers} we discuss the software solvers used in our evaluations, as well as notable solvers that were not suitable.  In Section \ref{sec:optimization} we present our experimental results on optimization.  In Section \ref{sec:sampling} we present our experimental results on sampling from ground states.  In Section \ref{sec:conclusion} we provide further discussion and conclude the paper.

\section{\dwave{} quantum processing units}\label{sec:dwave}

We start with an overview of  \dwave{} design features and introduce notation that will be used throughout.   For details about underlying technologies see \hiderefname{Bunyk et al.~}\cite{bunyk2014architectural}, \hiderefname{Dickson et al.~}\cite{Dickson2013}, \hiderefname{Harris et al.~}\cite{Harris2010},  \hiderefname{Johnson et al.~}\cite{johnson2011quantum} or \hiderefname{Lanting et al.~}\cite{Lanting2014}.

\subsection{Ising minimization}
\dwave{} annealing-based quantum processors are designed to find minimum-cost solutions to the  
Ising minimization (IM) problem,  defined  on a graph $G = (V,E)$ as follows.  Given a collection of fields $h = \{ h_i: i \in V \}$ and couplings $J = \{J_{ij}: (i, j) \in E \}$, assign values from $\{ -1, +1 \}$  to  $n$ {\em spin variables}  $s =\{ s_i \}$ 
so as to minimize the {\em energy function}         
\begin{eqnarray}\label{eqn:ising}
E( s ) &=  & \sum_{i\in V}  h_i  s_i   +  \sum_{(i,j)\in E} J_{ij} s_i  s_j.  
\end{eqnarray}
The spin variables $s$  can be interpreted as magnetic poles in a physical particle system;  in this context,  negative $J_{ij}$ is {\em ferromagnetic} 
and positive $J_{ij}$ is {\em antiferromagnetic},  the optimal solution is called a {\em ground state},  and nonoptimal solutions are {\em excited states}.  
IM instances can be trivially transformed to Quadratic Unconstrained Boolean Optimization (QUBO) instances defined on integers $s = \{0,1\}$,
or to Maximum Weighted 2-Satisfiability (MAX W2SAT) instances defined on Booleans $s=\{\text{true}, \text{false}\}$, all of which are 
NP-hard.   

\subsection{Chimera topology} 
The native connectivity topology for the \dwave{} QPU is based on a $C_{16}$ {\em Chimera graph} containing  $2048$ vertices (qubits) and $6016$ edges (couplers).  

A  Chimera graph of size $C_s$ is an $s\times s$ grid of Chimera cells (also called unit tiles or unit cells), each containing a complete bipartite graph on 8 vertices (a $K_{4,4}$).  Each vertex is connected to its four neighbors inside the cell as well as two neighbors (north/south  or east/west) outside the cell: therefore every vertex has degree 6 excluding boundary vertices.

In this study, as in others, we vary the problem size using square subgraphs of the full graph, from size $C_4$  (128 vertices) up to $C_{16}$ (2048 vertices).  Note that the number of problem variables $n=8s^2$ grows quadratically with Chimera size.  The reason we measure algorithm performance as a function of the Chimera size and not the number of qubits is that problem difficulty tends to scale exponentially with the Chimera size, i.e., with the square root of the number of qubits, since the treewidth of a Chimera graph $C_s$ is linear in $s$ \cite{robertson1986graph,Boixo2013}.

Because the chip fabrication and trapped magnetic flux leave some small number of qubits unusable, each processor has a specific  {\em hardware working graph}  $H \subset$ $C_{16}$.  The qubit yield---the fraction of qubits that are operational---is typically around 98\% for the 2000-qubit \dwave{} system whereas 95\% was typical for the \dwave{} 2X.  The working graph used in this study has 2035 working qubits out of 2048.

\subsection{Quantum annealing} 
\dwave{} processors solve Ising problems by {\em quantum annealing} (QA) in the form proposed by \hiderefname{Kadowaki and Nishimori }\cite{Kadowaki1998}.   The QA algorithm is implemented in hardware using a framework of analog control devices to 
manipulate a collection of qubit states according to a time-dependent Hamiltonian shown below.     
\begin{eqnarray}\label{eqn:aqc}
{\cal H}(t)  &= & A(t)  \cdot {\cal H}_{init}   +B(t) \cdot {\cal H}_{prob}. 
\end{eqnarray} 
QA carries out a gradual transition in
time $t : 0 \rightarrow t_a$,  from an initial ground state in ${\cal H}_{init}$, to a state described 
by the {\em problem Hamiltonian} ${\cal H}_{prob} = \sum_{i} h_i \sigma^z_i + \sum_{ij} J_{ij} \sigma^z_i \sigma^z_j$.   The problem Hamiltonian   
matches the energy function \eqref{eqn:ising},  so that a ground state for ${\cal H}_{prob}$  is a minimum-cost solution to $E(s)$. 

QA is closely related to \emph{adiabatic quantum computing} (AQC).  The AQC model of computation was proposed by \hiderefname{Farhi et al.~}\cite{Farhi2001} who showed that if the transition is carried out slowly enough the algorithm will find a ground state (i.e., an optimal solution) with high probability.

Theoretical guarantees about solution times for quantum algorithms (found in \cite{Farhi2001})  assume that the computation takes place in an ideal closed system,  perfectly isolated from energy interference from ambient surroundings.  The 2000-qubit \dwave{} chip is housed in a highly shielded chamber and cooled to near absolute zero; nevertheless, as is the case with any real-world  quantum device,  it must suffer some amount of interference,  which has the general effect of reducing the probability of landing in a ground state.  Thus, theoretical guarantees on performance may not apply to these systems.  We consider any  \dwave{} QPU to be a {\em heuristic} solver,  which requires empirical approaches to performance analysis.  
	
The \dwave{} QPU studied here contains 2035 active qubits (quantum bits) and 5912  active couplers made of microscopic loops of niobium connected to a large and complex analog control system via an arrangement of Josephson Junctions.  Thermometry on the refrigerator of the \dwave{} QPU and fits of single qubit measurements to a thermodynamic model indicate that $T\lesssim\text{\SI{15}{\milli\kelvin}}$.  When cooled to temperatures below \SI{9.3}{\kelvin}, niobium becomes a superconductor and is capable of displaying quantum properties including {superposition},  {entanglement}, and {quantum tunneling}.  Because of these properties, the qubits on the chip behave as a quantum mechanical particle process that carries out a transition from initial state described by ${\cal H}_{init}$ to a problem state described by ${\cal H}_{prob}$ \cite{boixo2016computational,Dickson2013,Lanting2014}.
	
\subsection{Modeling performance} 
Given input instance $(h,J)$,  a \dwave{} computation involves the following steps.   
\begin{enumerate}
\item {\bf Program.}   Load $(h,J)$ onto the chip; denote the elapsed programming/initialization time $t_i$.   
\item {\bf Anneal.}   Carry out the QA algorithm.  Anneal time $t_a$ can be set by user to some value $\text{\SI{5}{\micro\second}} \leq t_a \leq \text{\SI{1000}{\micro\second}}$.  

\item {\bf Read.}   Record qubit states to obtain a solution; denote the elapsed readout time $t_r$.
\item {\bf Repeat.}  Repeat steps b and c $k$ times to obtain a sample of $k$ solutions.  
\end{enumerate} 
We define {\em sample time}  $t_s$ and  {\em total time} $T$ as follows: 
\begin{eqnarray}
t_s & = &(t_a + t_r) \\ \nonumber
T & = & t_i  +  k\, t_s.  
\end{eqnarray}

For the \dwave{} system studied in this paper, the median programming time $t_i$ is \text{\SI{9.5}{\milli\second}} and the median readout time $t_r$ is \text{\SI{123}{\micro\second}}.

In this study, both for software solvers and for the \dwave{} QPU, we typically report annealing time rather than total time.  Annealing time is the measure of the algorithm proper, and measuring total time often obscures trends in data.  Scaling plots are particularly susceptible to this because the overhead of programming time makes scaling---typically presented on a semilog plot---look totally flat except for an uptick at the very largest problem sizes.  Further, we are most interested in the future potential of \dwave{} QPUs, and we expect that programming time and readout time will be reduced to small fractions of their current values; minimum annealing times will similarly be reduced, allowing us better control over the algorithm parameters.
For reference, since many people will be interested in total wall clock time, rather than annealing time, a 1000 times speedup over software solvers in annealing time, typical for the \dwave{} QPU in this study, translates roughly to a 30 times speedup in total wall clock time including programming and readout.

System characteristics of \dwave{} QPUs such as yield can vary within a generation.  If we compare this specific 2000-qubit \dwave{} system to the specific \dwave{} 2X QPU studied in 2015 \cite{King2015TTT}, programming time has decreased by 20\%, readout is three times faster, and yield has improved from 95\% to 99\%.

\section{Frustrated Cluster Loop problems}\label{sec:fcl}

\subsection{Ruggedness and clusters}

\emph{Ruggedness} is a feature of certain optimization problems---more specifically their energy landscapes---characterized by tall energy barriers and many local optima \cite{weinberger1990correlated,vassilev2000information}.  Typically, rugged problems are harder to solve, particularly with Markov chain Monte Carlo (MCMC) methods \cite{kendall2005markov,janke2007rugged}.  In the late 1980s, when ruggedness was first being explored in the context of evolutionary biology and bio-inspired computing, Kauffman's NK model was put forward as a model with tunable ruggedness inspired by genetic fitness functions under varying degrees of \emph{epistasis}, or how many other genetic loci affect the fitness contribution of a given locus \cite{Kauffman1987,Kauffman1989,Weinberger1996}.  The tunable ruggedness of the NK model has proved very valuable in the study of optimization heuristics, particularly evolutionary algorithms \cite{pelikan2009performance}.

Closely related to ruggedness is the analysis of spin overlap, in which landscape features are inferred from the distribution of overlap of two random states sampled from the Boltzmann distribution \cite{Yucesoy2013,Katzgraber2015good}.  Tall, thin peaks in the spin overlap distribution tend to correspond to tall, thin energy barriers; the presence of these features correlates not only with ruggedness and classical hardness, but also with applicability of quantum annealing, since quantum tunneling is likely to be a useful computational resource in the presence of these tall, thin barriers.  \hiderefname{Zhu et al.~}\cite{Zhu2016} have used spin overlap features to predict whether a problem can be solved by QA more efficiently than by SA, showing promising preliminary results for optimization problems such as weighted partial MAX-2SAT, minimum vertex cover, satisfiability, graph partitioning, circuit fault diagnosis, and certain spin-glass instances \cite{Zhu2016,Katzgraber2016AQC}.  This work points to the potential of quantum annealers to have a place in portfolio solvers \cite{gomes2001algorithm} and hybrid algorithms running on heterogeneous computing systems alongside CPUs, GPUs, and other coprocessors \cite{venkatasubramanian2009tuned,grewe2011static}.

To induce ruggedness using tall, thin energy barriers, Denchev et al.~\cite{Denchev2016} used ferromagnetically coupled unit tiles as clusters.  Flipping such a cluster in the absence of fields or external couplings involves jumping over or tunneling through an energy barrier that is 16 Ising units high and has a width of 8 in Hamming space.  Denchev et al.~\cite{Denchev2016} actually go beyond using single-tile clusters and use two-tile gadgets studied previously by Boixo et al.~\cite{boixo2016computational}.  This gadget is made up of two clusters that form a deceptive trap to draw annealers into a local minimum using local fields; annealers must then go over or through an energy barrier to reach the gadget's ground state.

Instead of using two-cluster gadgets, we simply use single-cell ferromagnetic clusters to induce ruggedness, leaving us with a simpler problem class.

\subsection{FCL problem generation}

We create local ruggedness by treating unit cells of the Chimera graph as ferromagnetically-coupled clusters.  We create global frustration by joining these clusters together using a problem generated on the logical graph of clusters.  This creates an energy landscape that is macroscopically interesting and in which the clusters induce wells separated by tall, thin energy barriers.

The logical graph of clusters is a square lattice, with a logical $16\times 16$ lattice of clusters spanning the working graph of a 2000-qubit \dwave{} QPU.\footnote{Note that a $16\times 16$ logical lattice is significantly larger than the largest logical lattice, $4\times 4$, of the Google problems considered by Denchev et al.~\cite{Denchev2016}---their two-tile gadgets take up more space than our one-tile clusters and the \dwave{} 2X has a smaller working graph than the 2000-qubit \dwave{} system.  These larger logical graphs in the problems we consider mean that the spin-glass backbones of these problems are significantly more computationally challenging.}  The problems we generate on the logical graph are \emph{frustrated loop problems}, constraint satisfaction problems first used in the evaluation of \dwave{} QPUs by Hen et al.~\cite{Hen2015} and modified to allow precision limits by King et al.~\cite{King2015}.

We refer to the final inputs as \emph{frustrated cluster loop} (FCL) problems.  For a given Chimera graph $G_C$ that may or may not have missing qubits or couplers, an FCL problem is generated from three parameters, $\alpha$ (the clauses-to-variables ratio), $\rho$ (the range, or precision), and $R \geq \rho$ (the ruggedness) as follows:

\begin{enumerate}
	\item Define each unit cell as a \emph{logical spin} if it has no missing qubits or couplers.  Use $c(v)$ to denote the logical spin index corresponding to qubit $v$.
	\item Wherever all four couplers connecting two logical spins are present, define these couplers as a \emph{logical coupler}.
	\item Define the logical graph $G_L$ as the graph comprising the logical spins and logical couplers.
	\item Generate a range-bounded frustrated loop problem Hamiltonian $(h_L, J_L)$ on $G_L$ using parameters $\alpha$ and $\rho$ as per King et al.~\cite{King2015} (note that $h_L$ is the zero vector).
	\item Define the native Chimera Hamiltonian $(h_C, J_C)$ with $h_C$ as the zero vector and $J_C$ as:
	$$
	J_C(u, v) =\left\{ \begin{array}{cl}
	-1, & \text{if $c(u) = c(v)$} \\
	\frac{1}{R}\cdot J_L(c(u), c(v)), &  \text{otherwise.} 
	\end{array} \right.
	$$
\end{enumerate}

It is worth repeating that these Hamiltonians have no fields (i.e., $h_L$ and $h_C$ are both zero vectors).  Note also that in-tile couplings in $J_C$ are all $-1$ and inter-tile couplings take values in $$\left\{j/R ~|~ j \in \left\{-R, -R+1, \dots, R\right\}\right\}.$$  Since we ensure that $\rho \leq R$, and $\rho \geq |j|$ for any logical coupling $j$, all couplings in $J_C$ are in the range $[-1, 1]$

The logical frustrated loop problems may be disconnected and have multiple components; we reject such disconnected inputs at generation time.

While these problems are large enough to span the entire working graph of the latest \dwave{} QPUs, the repetition code inherent in logical couplers and spins makes them relatively robust to analog errors \cite{Pudenz2015}.

\subsection{Problem class parameters}

The FCL problem class has three parameters: the clauses-to-variables ratio $\alpha$, the range $\rho$, and the ruggedness $R$.  We would like to restrict our experiments to the most interesting region of the parameter space.

\begin{figure}
\centering
\input{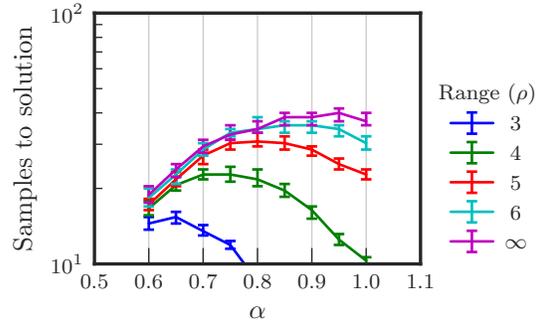}
\caption{Logical problem difficulty as measured by expected samples to solution for simulated annealing.  Error bars show the 95\% confidence intervals for the medians, grouped over $\alpha$ and $\rho$.  Difficulty is maximized at $\alpha=0.65$ for precision 3, $\alpha=0.75$ for precision 4, $\alpha=0.8$ for precision 5, and $\alpha=0.85$ for precision 6.\label{fig:alpha}}
\end{figure}

First we aim to determine the value of $\alpha$ that maximizes the difficulty of the logical problem.  If $\alpha$ is too low, a problem is underconstrained and is easy to solve.  If $\alpha$ is too high, the planted solution is expressed too strongly and the problem's features approach those of a ferromagnet, making it easy to solve.  The difficulty of the logical problem depends only on $\alpha$ and $\rho$, not on $R$.  For various values of $\rho$, we perform a sweep of $\alpha$ to determine the value that maximizes the hardness of the logical problem (see Figure \ref{fig:alpha}).

The impacts of $\rho$ are more nuanced.  First, the value of $\rho$ provides an upper bound on the limit of $\alpha$ because packing in more loops eventually raises the maximum coupler range.  Second, for a fixed value of $\alpha$, problems with a lower $\rho$ value have their loops spread out more evenly over the logical spins.  Finally, for the native problem, coupler values are scaled down by a factor of $R \geq \rho$ so that inter-tile couplings are in the range $[-1, 1]$ (in-tile couplings are always $-1$).  Thus higher values of $\rho$ constrain $R$ to be higher, and make problems more locally rugged relative to the global Hamiltonian.  In Section \ref{sec:optimization}, we attempt to deconvolve the impacts of $\rho$ and $R$.

The ability to tune the ruggedness of the inputs by varying $R$, either by specifying $R = \rho$ and varying $\rho$, or by varying $R$ independently, gives FCL problems an additional degree of utility, particularly when assessing the value of quantum tunneling and the potential of quantum annealing.  Varying $\rho$ and specifying $R = \rho$ makes problem generation simpler by reducing the number of free parameters whereas fixing $\rho$ and varying $R$ allows us to isolate the impact of ruggedness without altering the complexity of the logical problem.

\subsection{Confirming correlation between ruggedness and classical hardness}

We expect to see a positive correlation between ruggedness and classical hardness.  Here we characterize classical hardness using the decorrelation time of an MCMC procedure.  To validate this assumption we measure the decorrelation time for a parallel tempering (PT) procedure that uses the Metropolis algorithm in combination with the standard replica exchange rule~\cite{Hukushima:PT}; we use the autocorrelation of temperature as our measure of decorrelation~\cite{Banos:NSG2010}.  For more details of this method, see Appendix \ref{app:decorrelation}.

\begin{figure}
	\centering
	\input{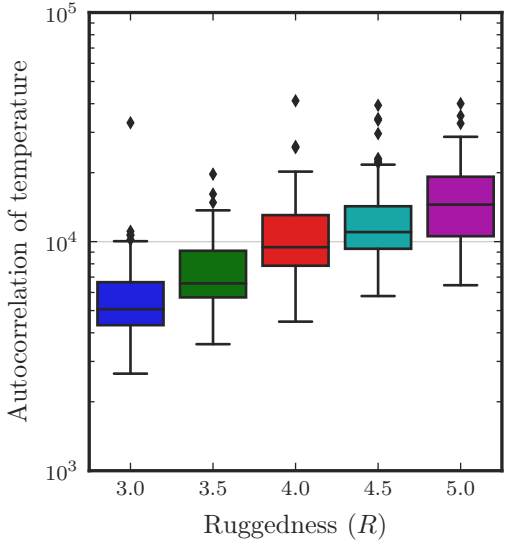}
	\caption{Box plots showing ruggedness versus classical hardness.  We hold $\rho$ and $\alpha$ fixed at 3 and 0.65, respectively, and vary the ruggedness $R$ of the native Chimera problem.  At each value of $R$ we generated 100 instances; points indicate outliers.  Classical hardness is measured using the autocorrelation of temperature.  There is a clear positive correlation between ruggedness and classical hardness.\label{fig:decorrelation}}
\end{figure}

Figure \ref{fig:decorrelation} illustrates the relationship between ruggedness and classical hardness.  Confirming our intuition, FCL problems with greater ruggedness are characterized by greater classical hardness.

\section{Software solvers}\label{sec:solvers}

The four software solvers we consider are GPU implementations of SA, QMC, and SVMC, and Selby's CPU implementation of HFS \cite{Selby2014,Selby2013git}.  

Recent studies of \dwave{} QPUs have not included GPU-based software solvers despite the fact that SA is very amenable to GPU implementation \cite{Ferreiro2012}.  The addition of GPU solvers is a significant raising of the bar in terms of software competition, and means that solvers that can be implemented on GPUs have taken a leap forward relative to solvers that cannot.  Run on modern hardware, our GPU-based algorithm implementations are roughly 1000 times faster than the corresponding single-core CPU implementations.

Mandr\`{a} et al.~\cite{Mandra2016} analyzed the performance of a diverse set of solvers on the inputs of Denchev et al.~\cite{Denchev2016}. However the lack of GPU implementations of these solvers means that most are unlikely to be competitive in an absolute sense.  Indeed, of the three classes of solvers that they study, only sequential algorithms, which they find to have the worst performance, have the massive parallelizability and low memory requirements that make efficient GPU implementations possible.  It is also possible to implement SA in a field-programmable gate array (FPGA), but the additional speedup over GPU implementation is limited and generally not worth the increased cost of hardware.

In Appendix \ref{sec:solvers-detail} we give further details of the software solvers and parameterizations we used.  We also discuss algorithms we omitted because of prohibitive runtimes.

\section{Optimization}\label{sec:optimization}

We measure the expected time to solution (TTS) of different solvers on the inputs, calculated as \[\text{TTS} = \frac{\text{time per anneal}}{\text{ground state probability}}.\]  We consider only annealing times and exclude programming and readout times from our analysis as these are not part of the algorithms proper.

For a given value of $\rho$, we choose $\alpha$ to maximize the difficulty of the logical problem (see Figure \ref{fig:alpha}).  For each selection of $\rho$ and $R$, we generate 100 FCL problems at each problem size and solve each problem with each solver.

\begin{figure*}
	\centering
	\input{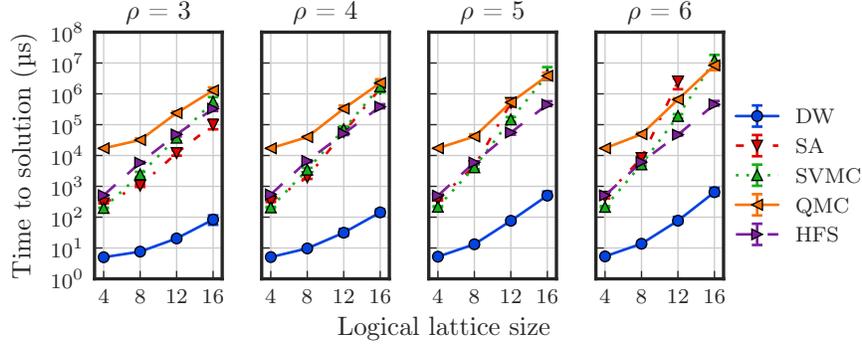}
	\caption{Time to solution for \dwave{} and software solvers with range values $\rho\in\{3, 4, 5, 6\}$.  For each value of $\rho$, $\alpha$ is chosen to maximize logical hardness.  Shown are median values (over 100 inputs at each size) with 95\% confidence intervals.\label{fig:tts}}
\end{figure*}

\subsection{Varying ruggedness via logical complexity}

In our first experiment, we vary $\rho$ and set $R = \rho$.  In this case the only free parameter $\rho$ controls both the ruggedness and the logical complexity of the inputs.  Time-to-solution plots are shown in Figure \ref{fig:tts}.  At the largest problem size, the \dwave{} QPU is three orders of magnitude faster than the fastest software solver for each value of $\rho$.  \dwave{}'s speedup over software peaks at 2600 times for $\rho=4$.

As $\rho$ increases, the impact of local ruggedness increases as the logical Hamiltonian is compressed relative to the local wells induced by the clusters.  The performance of SA drops off sharply while the performance of DW and QMC declines gracefully.  The performance of HFS decreases only very slightly.  HFS is not affected by the local ruggedness because it is tailored to the Chimera topology and uses updates that contain entire clusters; the performance degradation is due to the slight increase in logical problem hardness.
	
All solvers except HFS have strictly convex scaling curves because the anneal lengths are optimized for the largest problem size and are too long for the smaller problems.  HFS does not use fixed-length anneals and ends up using shorter anneals on smaller inputs.

Though true scaling is masked by the inability to optimize parameters for smaller inputs \cite{Ronnow2014,Amin2015}, we note that the performance of the \dwave{} QPU scales at least as well as the software solvers between the two largest problem sizes.

\subsection{Varying ruggedness by scaling}

In our second experiment, we fix $\rho=3$ and vary the ruggedness $R$.  This keeps the logical complexity constant, allowing us to isolate the impact of ruggedness on the various solvers.  Here we consider only the largest problem size having a $16\times 16$ logical lattice.

\begin{figure}
	\centering
	\input{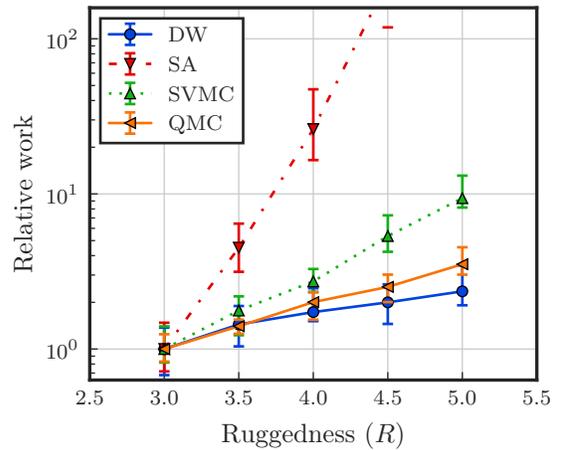}
	\caption{Ruggedness (increasing from left to right) versus relative work for various solvers.  Relative work for each solver is calculated as TTS divided by median TTS at $R=3.0$.  The solvers have notably different responses to increasing ruggedness, with SA struggling the most, followed by SVMC, then QMC, then the \dwave{} QPU.  HFS deals with these energy barriers using exponential brute force; therefore the parameter $R$ does not affect its performance.  Markers indicate medians (over 100 inputs) and error bars indicate 95\% confidence intervals for the median.\label{fig:work_vs_ruggedness}}
\end{figure}

Consistent with our findings when varying $\rho$, tuning the ruggedness directly by varying $R$ increases difficulty dramatically for simulated annealing and less so for other solvers (see Figure \ref{fig:work_vs_ruggedness}).  Excluding HFS, whose behavior is constant in this example, the work required by a solver essentially scales according to its quantumness.  The \dwave{} QPU is most capable of dealing with ruggedness.  QMC---the most faithful classical simulation of quantum annealing---comes next, followed by SVMC, which is a mean-field approximation to QMC.  Bringing up the rear is SA, a simulation of a fully classical process.

The improved scaling (versus ruggedness) of QMC over SVMC indicates that crucial information is being lost in the mean-field approximation.  The improved scaling of QA (i.e., the \dwave{} QPU) over QMC may indicate that QMC is failing to faithfully simulate the dynamics of the QA processor, or it may simply be an artifact of our inability to use faster \dwave{} anneals.  This bears further investigation using future \dwave{} QPUs with faster annealing times, again utilizing the tunable ruggedness of FCL problems.

\section{Sampling}\label{sec:sampling}

The ability of an Ising solver to sample diverse optima has both practical and theoretical importance.  Ground state sampling in combinatorial problems is the basis for construction of space-efficient SAT-based membership filters \cite{Weaver2014,Douglass2015}.  The associated complexity class, \#P ---the counting analog of NP---has been the subject of extensive research in theoretical computer science since the 1970s \cite{valiant1979complexity}.  Sampling from the Boltzmann distribution, in which  states with equal energy are sampled with equal probability, is of particular interest in machine learning.  Boltzmann samples are used to train Boltzmann machines, a task known to be both hard and useful \cite{long2010restricted}.

While machine learning applications typically depend on finite-temperature Boltzmann sampling, using near-optimal states as well as optimal states, we focus on zero temperature sampling to simplify our investigation.  This saves us from having an additional input parameter $\beta$---the inverse temperature---that we would have to either set arbitrarily or determine empirically.  Empirical estimation of $\beta$ can be challenging \cite{Raymond2016} and basing the target $\beta$ on the output of a solver would arguably give that solver an unfair advantage.  

\subsection{Sampling from all valleys}

The expected time required for a solver to find all ground states of a problem is known, both in the equiprobable case and the biased case \cite{Flajolet1992}.  In the case of an Ising spin problem, ground states often lie in connected valleys in Hamming space, and given one ground state in the cluster it is easy to find the rest.  We therefore adopt a more practical metric based on the time required to sample all {\em valleys} of ground states.

In ground states of FCL problems, all clusters have their spins in agreement; therefore the distance between any two ground states in the native Hamming space is a multiple of 8.  However, ground states can be adjacent (i.e., differ by a single spin) in the logical space.  We define a \emph{valley} as a set of ground states that are connected in the logical Hamming space.  While it is nontrivial to move from one state to another in the same valley because of the single tall, thin energy barrier, it can still be done with a modest amount of postprocessing.  We also note that, since FCL problems do not have fields, ground states come in antipodal pairs,\footnote{For Hamiltonians with no fields, flipping all spins of a state does not change the energy.  Therefore the antipode (negation) of any ground state is also a ground state.} and by extension so do valleys.  We treat each pair of antipodal valleys as a single valley since it is trivial to move from one to the other.

With valleys defined in this way, we define the time-to-all-valleys (TTAV) metric as the expected amount of annealing time required to draw at least one sample from each valley.  This metric captures the hardest part of sampling from these distributions---finding ground states in every mode---and ensures a diverse set of solutions.  With at least one sample from each valley, it is possible to find all ground states using only a modest amount of postprocessing.  The TTAV metric is most meaningfully interpreted relative to TTS since hitting valleys directly depends on hitting ground states.

\subsection{Mining for interesting valley structure}

Sampling from all valleys is not always much harder than finding a single ground state---an input may have only a single valley or may have valleys that are all very close in Hamming space.  We wish to generate inputs with multiple valleys that are well-separated.  Sampling from distributions with multiple, well-separated valleys is particularly hard \cite{Neal1996} and has important applications such as classification using deep Boltzmann machines \cite{srivastava2012multimodal}.

Because we define valleys as clusters of ground states in the \emph{logical} space, analyzing the valley structure of an input is tractable.  The largest logical graphs are $16\times 16$ lattices having treewidth 16, so solving a logical problem using dynamic programming and returning some fixed number of ground states typically takes less than a second.  This allows us to mine for inputs having interesting valley structure.

We quantify interesting valley structure using the distribution of spin overlap $P(q)$ \cite{Yucesoy2013,Katzgraber2015good} at zero temperature, i.e., for two ground states sampled uniformly with replacement, what fraction of spins do they have in common?  The random variable $P(q)$ takes values in the range $[-1,1]$.  For inputs without fields the distribution is symmetric about zero; we can therefore consider the distribution of the absolute value $P(|q|)$.  We define the \emph{mean overlap} as the expectation of $P(|q|)$.  Inputs with mean overlap near 1 tend to resemble ferromagnets---if there are multiple valleys they will be close together.  Inputs with lower mean overlap tend to have valleys that are well-separated.

Inputs that are hard to sample from have multiple valleys that are well-separated.  We mine for such inputs as follows.  First we reject any input with more than 1000 ground states, as these slow down our analysis and may be too easy.  Second, we reject any input that does not have at least 4 valleys since we want valley collection to be nontrivial.  Finally, we reject any input with a mean overlap of 0.7 or higher since we want valleys to be well-separated.  

\subsection{Sampling results}

We generated problems at the $16\times 16$ lattice size with $\alpha=0.85$ and $\rho=R=6$ and mined them for interesting valley structure as described above.  We generated 50,000 inputs and rejected all but 74.  This gave us an acceptance rate of roughly $0.15\%$ of inputs.  We sought to answer the question, after $x$ seconds of annealing, what fraction of the valleys has each solver seen?  For each problem we drew a number of samples according to the solver as follows:

\begin{center}\begin{small}
\begin{tabular}{lp{5cm}}
\vspace{1mm}\dwave{} QPU:& 100,000 samples at \SI{5}{\micro\second} (in batches of 100 per spin-reversal transform) \\
\vspace{1mm}SA: &100,000 samples \\
\vspace{1mm}QMC: &5000 samples \\
HFS: &5000 samples
\end{tabular}\end{small}\vspace{-1em}
\end{center}

\begin{figure*}[t!]
	\centering
	\input{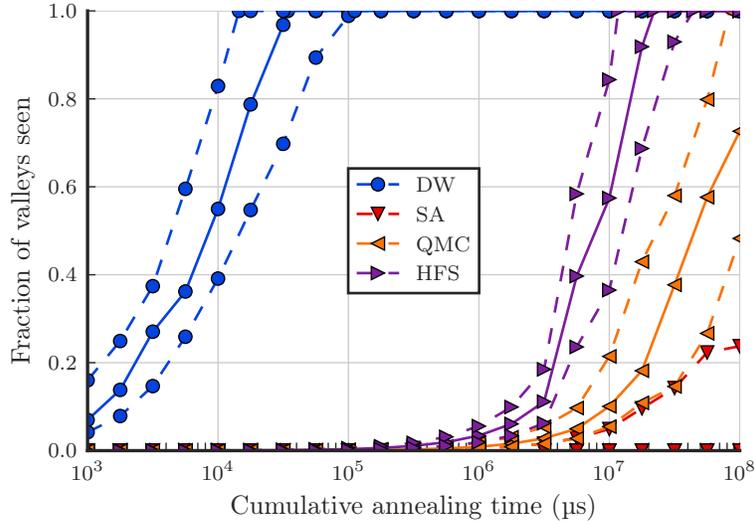}
	\caption{Time to all valleys (TTAV) for various solvers.  The $x$-axis shows elapsed annealing time and the $y$-axis shows the fraction of valleys that a solver has hit up to that point in time.  Solid lines show medians (over 74 inputs) and dashed lines show the 25th and 75th percentiles.\label{fig:ttav}}
\end{figure*}

\subsubsection{Time to all valleys}

Results for the TTAV metric are shown in Figure \ref{fig:ttav}.  The 2000-qubit \dwave{} QPU is the fastest of the solvers, hitting all valleys in a median time of roughly \SI{30}{\milli\second}.  The fastest software competition was HFS, which hit all valleys in a median time of roughly \SI{30}{\second}.  

In certain situations quantum annealing in the transverse-field Ising model is subject to inherent sampling bias \cite{Albash2015PRA,Matsuda2009,Mandra2016bias,king2016floppy}, although that does not prove to be a significant problem here.  While the slope of the \dwave{} curve is slightly less steep than the HFS curve, indicating that its samples might be less diverse, the \dwave{} QPU still manages to outperform the competition by about three orders of magnitude.

\subsubsection{KL-divergence of valley distributions}

The TTAV results shown in Figure \ref{fig:ttav} fail to address a specific fear---that in a significant minority of inputs there are valleys that the \dwave{} QPU would be simply unable to find due to quantum sampling bias.  To address this, we calculate for each (solver, problem) pair the KL-divergence between empirical valley distributions and exact valley distributions (i.e., relative valley sizes).  KL-divergence is an \emph{asymmetric} measure of the distance between two probability distributions; we calculate it such that it is infinite if the solver fails to see all valleys, i.e., 
\[ \text{KLD}=\sum_{\text{valleys~} v} P(v)\log{\frac{P(v)}{\widehat{P}(v)}}\,, \]
where $P(v)$ is the true Boltzmann probability of valley $v$ and $\widehat{P}(v)$ is the sample estimate of $P(v)$, conditioned on samples being ground states.
This KL-divergence measure includes two types of error.  First, there is a distributional error, since each solver samples from a distribution that differs from the Boltzmann distribution.  Second, there is a sample size error, since our sample estimate has finite size and therefore differs from the solver's true distribution.  In this context it is appropriate to include both types of error.

\begin{figure*}[t!]
	\centering
	\input{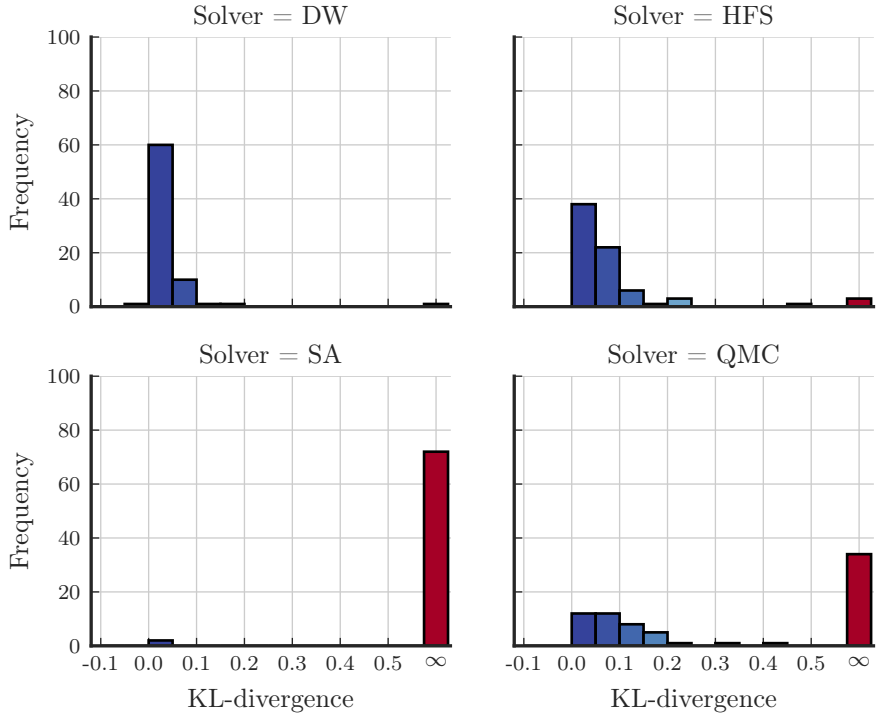}
	\caption{KL-divergence histograms.  Shown are the empirical distributions (out of 74 inputs) of the KL-divergence achieved by each solver in estimating the valley distributions.  Where the value is infinite, the solver failed to see one or more of the valleys.  The \dwave{} QPU had the best performance in this metric---even with three orders of magnitude less annealing time---followed by HFS, then QMC, then SA.\label{fig:kld_histograms}}
\end{figure*}

Figure \ref{fig:kld_histograms} shows histograms of KL-divergence for the different solvers.  For these FCL problems, fears of valleys suppressed by quantum sampling bias are unfounded.  The \dwave{} QPU has a superior KL-divergence distribution than any of the software solvers even when annealing for three orders of magnitude less time.  On the single input for which the \dwave{} KLD was infinite because at least one valley was never seen, it was also infinite for all other solvers.

\subsubsection{Raw error on model marginals}

The TTAV metric and valley distributions can be thought of as representing what sample quality would look like with postprocessing.  We would also like a more raw metric that does not have this implicit postprocessing.  For this we consider marginals of the zero-temperature Boltzmann distribution.  Specifically, we consider the spin-spin expectations, i.e., for each coupler, what is the expected product of the two incident spins in the zero-temperature Boltzmann distribution?  The $L_1$ error on these marginals (i.e., the empirical estimates of spin-spin expectations minus true expectations) is a well-established metric of interest in the study of undirected graphical models (see, e.g., \cite{Globerson2007}).

\begin{figure*}
	\centering
	\input{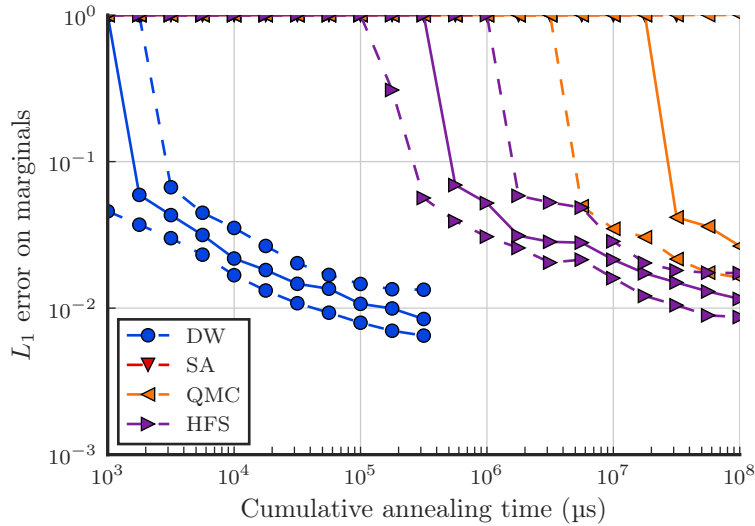}
	\caption{Elapsed annealing time versus $L_1$ error of marginal estimation for various solvers.  Solid lines show medians (over 74 inputs) and dashed lines show the 25th and 75th percentiles.  The \dwave{} QPU achieves the same error as software solvers in roughly three orders of magnitude less time.\label{fig:l1_error}}
\end{figure*}

Figure \ref{fig:l1_error} shows the decay in error as more samples are taken.  We measure errors in the logical space, so couplers within a cluster are ignored.  Again, the \dwave{} QPU is roughly three orders of magnitude faster than the fastest software solver.  We expect that the error on marginals will decay to a certain point, then plateau, with the level of the plateau corresponding to the bias of the solver.  As with the TTAV data, potential concerns about the bias of quantum annealers do not seem to play out here.  The $L_1$ error of the \dwave{} QPU is still decaying after 100,000 samples; at this point in time (\SI{0.5}{\second}) the \dwave{} QPU has achieved a median error of less than 1\%, a value that software solvers fail to reach in \SI{100}{\second} to match.

\section{Conclusions}\label{sec:conclusion}

We have introduced a class of synthetic inputs on which to evaluate the performance of annealing-based QPUs, specifically the QPUs developed by D-Wave.  This problem class is more representative of real-world problems and provides an alternative to the Google problems of Denchev et al.~\cite{Denchev2016}, which were more highly tuned to highlight the utility of quantum tunneling.

The \dwave{} QPU is up to 2600 times faster than all software solvers considered and typically on the order of 1000 times faster at the largest problem size.  These software solvers now include GPU implementations of SA, QMC, and SVMC as well as a CPU implementation of HFS, making the competition much stronger than that analyzed by Denchev et al.~\cite{Denchev2016}.  The set of software solvers we used was representative, but not exhaustive---in particular, we hope to include more of the solvers used by Mandr\`{a} et al.~\cite{Mandra2016} in future studies.

These inputs have tunable ruggedness controlled either by the range parameter $\rho$ or by the scaling parameter $S$.  Of particular interest is the fact that QMC performed better relative to SA when the ruggedness is increased, and physical quantum annealing performed better still.  

We also evaluated the 2000-qubit \dwave{} QPU on the task of zero-temperature Boltzmann sampling, i.e., sampling uniformly from ground states.  While concerns have been raised that quantum and analog sampling bias might make it difficult for quantum annealers to sample from Boltzmann distributions \cite{Albash2015PRA,Matsuda2009,Mandra2016bias}, there was little evidence for such a struggle in this study.  In several metrics considered, the 2000-qubit \dwave{} QPU maintains its speed advantage of roughly three orders of magnitude and provides sample diversity that is as good as, or better than, the software competition.

\bibliography{paper}
%\printbibliography

\appendix

\section{Calculation of decorrelation}\label{app:decorrelation}

The parallel tempering implementation we use to measure decorrelation time is parameterized by a sequence of $n$ increasing inverse temperatures $\beta_i \in [0,\beta=3]$.  A replica sample is initialized randomly for each temperature.  The replicas are then iterated, undertaking a random walk in temperature space combined with MCMC sweeps controlled by the energy landscape\footnote{At $\beta=0$, the hottest replica, we replace the sample with a new random uniformly drawn sample on each iteration so that decorrelation is perfect.}. Inference on the lower temperature distributions is hardest. The quality of inference is limited by the time scale associated with the temperature random walk.  For a sample to decorrelate at low temperature on a practical time scale, the random walk must pass through a high temperature state (where decorrelation is fast) and return to the low temperature state. The time scale is approximated by the integrated autocorrelation of the temperature index~\cite{Banos:NSG2010}. 

In our setup, temperatures are selected independently for every instance, with $\beta$ spanning $[0,30]$ such that the replica exchange rate is equalized at close to 40\% (no lower than 33\%, no higher than 50\%). Equalization of exchange rates is an intuitive and well-studied heuristic for temperature selection and is optimal in special cases~\cite{Kone:ST}.\footnote{We note that, as would be true in studying any problem class, better choices for temperatures and transition operators can lower the autocorrelation times relative to those presented.} Equalization of exchange rates is achieved heuristically by iterating PT, refining the temperature set by linear interpolation of the log empirical exchange rates. To measure autocorrelation time, we undertake a long run of 600,000 sweeps, discarding a conservative portion (10\%)\hide{(io: so far I've been discarding half of it since the autocorrelation times did not change from 10\% and 50\% - I can easily calculate for 10\% if necessary, kept all the data)} of the initial samples which we took as sufficient for burn in (this assumption was tested self-consistently). We then extract the integrated autocorrelation time from the empirical values by an initial sequence estimator~\cite{Janke:SA}, we average over the autocorrelations on the $n$ available chains to reduce noise.

\section{Details of software solvers}\label{sec:solvers-detail}

\subsection{Classical hardware}

\begin{table*}
	\begin{center}\begin{small}
			\begin{tabular}{|c|c|c|}
				\hline
				{\bf Resource} & CPU & GPU \\ \hline
				{\bf Model} & \specialcell{Intel\reg Xeon\reg CPU\vspace{-1mm}\\E5-2643 v3}  & \specialcell{NVIDIA\reg GeForce\reg\vspace{-1mm}\\GTX 1080} \\
				{\bf Clock rate} & \SI{3.4}{\giga\hertz} & \SI{1.6}{\giga\hertz} \\
				{\bf Cores } & 6 & 2560 \\
				{\bf Concurrent workers } & 6 & 1 \\
				{\bf Power} & \SI{135}{\watt} & \SI{180}{\watt} \\ \hline
			\end{tabular}\end{small}\vspace{-1em}
		\end{center}\caption{Specifications for classical processors used for software solvers.\label{table:processors}}
	\end{table*}

	Table \ref{table:processors} contains the specifications for the classical processors used.  CPU algorithms were run single-threaded on one core each; multiple workers used cores concurrently for independent jobs.  GPU algorithms are highly parallelized and each GPU job uses the entire GPU.

\subsection{Included software solvers}

Our experiments and analyses focus on four algorithms commonly used in comparisons with \dwave{} QPUs---three that are highly amenable to GPU implementation and one that is highly tailored to the Chimera topology.

\subsubsection{Simulated annealing}
Simulated annealing \cite{Kirkpatrick1983} is a simulation of thermal annealing that is widely used as an optimization algorithm.  It is the classical analog to quantum annealing.  Since simulated annealing is a simple algorithm with very low memory requirements and a high degree of parallelizability, it is ideal for implementation on a GPU.

\subsubsection{Quantum Monte Carlo}
Quantum Monte Carlo, also known as simulated quantum annealing, is a classical approximation to quantum annealing.  For the algorithm to work efficiently on a GPU, we implement the discrete time variant of QMC and fix the number of Trotter slices at 64 so that a worldline can be packed as bits in a word.

While the continuous time variant of QMC is a more faithful simulation of quantum annealing, in particular serving as a bias-free sampler that approaches the quantum Boltzmann distribution in the limit, it has been shown that discrete time QMC can have superior performance as an optimizer \cite{Heim2015}.

\subsubsection{Spin vector Monte Carlo}
Spin vector Monte Carlo (SVMC), also known as the O(2)-rotor model, is a mean-field approximation to QMC.  SVMC can be thought of as falling between SA and QMC.  Proposed for use as an approximation to \dwave{} QPUs by Shin, Smith, Smolin, and Vazirani \cite{Shin2014}, it is also known as the SSSV algorithm.  We use a GPU implementation of SVMC that is a minor modification of our implementation of SA.

\subsubsection{Hamze-de Freitas-Selby}
The Hamze-de Freitas-Selby (HFS) algorithm optimizes by repeatedly optimizing the spins in a low-treewidth induced subgraph of the input, subject to the rest of the input being fixed.  The subgraph over which the input is optimized changes at each step.  The HFS algorithm is a greedy search algorithm in which reassignment of many variables is considered at once.  We used Selby's implementation \cite{Selby2014,Selby2013git} that is heavily tailored to the Chimera topology; we modified this solver to return each stopping state for consistency with the other heuristic solvers.  We note that the HFS algorithm cannot be efficiently implemented on GPU because the memory requirements are too high.

\subsection{Excluded software solvers}

In addition to these four algorithms, we considered several other software solvers that were prohibitively slow; due to limited time and resources it was not feasible to perform the long software runs needed to optimize parameters and determine ideal performance.

\subsubsection{Nontailored HFS}
We tested an implementation of HFS that, rather than using subgraphs tailored to the Chimera topology, is topology-agnostic and generates subgraphs dynamically.  This nontailored version of HFS performed far worse than Selby's tailored implementation, to the point where we failed to hit ground states in the largest problems.  The failure of this algorithm highlights the extent to which Selby's HFS implementation, and specifically the hardcoded subgraphs to update, exploit the sparsity and modularity of the Chimera topology \cite{Mandra2016}.  It is very likely that this type of exploitation will be impossible in future quantum annealer topologies \cite{Denchev2016}.

\subsubsection{Wolff cluster Monte Carlo}
The Wolff algorithm \cite{Wolff1989} dynamically detects clusters of spins that should be flipped together.  We used a modified implementation that considers the potential change in energy when deciding whether to flip a cluster, similar to Venturelli et al.~\cite{Venturelli2014}.  This algorithm would, at first glance, be ideal for FCL problems due to the crucial role of clusters.  However, finding clusters is slow and our CPU implementation was not competitive with other solvers.  Note that the Wolff algorithm is not particularly amenable to GPU implementation; such implementations exist for topologies such as lattices \cite{Komura2012} but they achieve only modest speedups over CPU implementations.

\subsubsection{Parallel tempering}
Parallel tempering runs multiple replicas of a Monte Carlo simulation at different temperatures in parallel, and can exchange information between replicas according to certain exchange rules.  It is the algorithm of choice for approximately calculating features and statistics of an energy landscape when exact calculation is prohibitive.  Our GPU implementation used 64 replicas, with replicas of a spin packed bitwise into a word, similar to our implementation of QMC.  Parallel tempering is more powerful than simulated annealing, but in this case proved to be uncompetitive due to the increased cost of each step.

\subsubsection{PT-ICM}
In vanilla parallel tempering, replica exchange steps are combined with single-spin Monte Carlo updates.  However, replica exchanges can be combined with other types of updates.  Isoenergetic cluster move (ICM) updates can be combined with replica exchange and simple Monte Carlo updates.  This variant was introduced by Swendsen and Wang \cite{swendsen1986replica} in their original parallel tempering paper and is sometimes called the PT-ICM algorithm.  We implemented PT-ICM as described by Zhu et al.~\cite{Zhu2016borealis} and found that its scaling was similar to Selby's implementation of HFS, but absolute performance was an order of magnitude slower even when using a precomputed ideal temperature ladder (i.e., set of $\beta$ values) specific to each input.

The advantage that PT-ICM has over Selby's implementation of HFS is that PT-ICM detects clusters dynamically and is not tailored to the underlying topology.  Because of this, there may be a misconception that PT-ICM will be future proof against denser quantum processor topologies.  However, problems on denser topologies will have smaller site-percolation thresholds \cite{bollobas2006percolation}, and the effectiveness of PT-ICM depends crucially on a large site-percolation threshold \cite{Zhu2016borealis}.  Thus increased processor density will erode the computational value of cluster updates and consequently the advantage of PT-ICM over vanilla PT.

\subsection{Parameter tuning}

For the \dwave{} QPU, optimal performance at all problem sizes was achieved at the minimum allowed annealing time of \SI{5}{\micro\second}.  It therefore makes sense to optimize the parameters of software solvers only at the largest problem size; optimizing on a per-size basis would only make scaling look worse for the software solvers.  Since we cannot properly optimize the performance of all solvers at all problem sizes, we focus on results at the largest problem size and take scaling results with a grain of salt.

For SA, we chose values of $\beta$ that increase linearly from 0.01 to 3 and used $10^5$ sweeps.  For QMC, we used a fixed $\beta$ of 30 and $10^4$ sweeps with the transverse field $A(t)$ decreasing linearly from 1 to 0 and the longitudinal field $B(t)$ increasing linearly from 0 to 1.  For SVMC we again used a $\beta$ of 30 and the same annealing schedule, but used $10^5$ sweeps.  These values of $\beta$ were chosen to optimize performance.  For HFS, we used Selby's strategy GS-TW2 \cite{Selby2014}.  For the \dwave{} QPU, we use the minimum annealing time of \SI{5}{\micro\second}.

%%%%%%%%%%%%%%%%%%%%%%%%%%%%%%%%%%%%%%%%%%%%%%%%%%%%%%%%%%%%%%%%%%%%%%%%%
%%%%%%%%%%%%%%%%%%%%%%%%%%%%%%%%%%%%%%%%%%%%%%%%%%%%%%%%%%%%%%%%%%%%%%%%%
%%
%% ENDMATTER
%%
%%%%%%%%%%%%%%%%%%%%%%%%%%%%%%%%%%%%%%%%%%%%%%%%%%%%%%%%%%%%%%%%%%%%%%%%%
%%%%%%%%%%%%%%%%%%%%%%%%%%%%%%%%%%%%%%%%%%%%%%%%%%%%%%%%%%%%%%%%%%%%%%%%%

% End the document.
\end{document}